\documentclass[12pt]{article}
\usepackage{amssymb,amsmath,fullpage}
\usepackage{graphicx}

\title{On the Sensitivity of\\
Cyclically-Invariant Boolean Functions}
\author{Sourav Chakraborty\\University of Chicago\\
\texttt{sourav@cs.uchicago.edu}}
\date{05-10-2004}

\newtheorem{theorem}{Theorem}[section]
\newtheorem{lemma}[theorem]{Lemma}
\newtheorem{proposition}[theorem]{Proposition}
\newtheorem{claim}[theorem]{Claim}
\newtheorem{corollary}[theorem]{Corollary}
\newtheorem{example}[theorem]{Example}

\newtheorem{definition}[theorem]{Definition}
\newenvironment{proof}{{\bf Proof:}}{\hfill\rule{2mm}{2mm}}

\newenvironment{clmproof}{{\bf Proof of Claim:}}{\hfill\rule{2mm}{2mm}}
\newenvironment{propproof}{{\bf Proof of Proposition~\ref{pro}:}}{\hfill\rule{2mm}{2mm}}
\newenvironment{thmproof1}{{\bf Proof of
Theorem ~\ref{thm1}:}}{\hfill\rule{2mm}{2mm}}

\def\bool{\{0,1\}}
\def\ie{i.\,e.}

\DeclareMathOperator{\wt}{wt}
\DeclareMathOperator{\supp}{supp}

\begin{document}
\maketitle

\begin{abstract}

In this paper we construct a cyclically invariant Boolean function
whose sensitivity is $\Theta(n^{1/3})$. This result answers two
previously published questions. Tur\'an (1984) asked if any
Boolean function, invariant under some transitive group of
permutations, has sensitivity $\Omega(\sqrt{n})$. Kenyon and Kutin
(2004) asked whether for a ``nice'' function the product of
0-sensitivity and 1-sensitivity is $\Omega(n)$. Our function
answers both questions in the negative.

We also prove that for minterm-transitive functions (a natural class of
Boolean functions including our example) the sensitivity is $\Omega(n^{1/3})$.
Hence for this class of functions sensitivity and block
sensitivity are polynomially related.

\end{abstract}

\section{Introduction}
Cook, Dwork and Reischuk \cite{cook-dwork-reischuk-86} originally
introduced sensitivity as a simple combinatorial complexity
measure for Boolean functions providing lower bounds on the time
needed by a CREW PRAM. Nisan \cite{nisan-91-1} introduced the
concept of block sensitivity and demonstrated the remarkable fact
that block sensitivity and CREW PRAM complexity are polynomially related.
Whether block sensitivity and sensitivity are polynomially related is still
an open question.

The largest known gap between them is quadratic, as shown by
Rubinstein \cite{rubinstein-95}. But for an arbitrary Boolean
function the best known upper bound on block sensitivity in terms
of sensitivity is exponential. H.-U. Simon \cite{h.u.simon-83}
gave the best possible lower bound on sensitivity in terms of the
number of effective variables. From that it follows that block
sensitivity of a function $f$ is $O(s(f)4^{s(f)})$, where $s(f)$
is the sensitivity of the function $f$. Kenyon and Kutin
\cite{kutin-kenyon-01} gave the best known upper bound on block
sensitivity in terms of sensitivity; their bound is
$O\left(\frac{e}{\sqrt{2\pi}}e^{s(f)}\sqrt{s(f)}\right)$.

Nisan pointed out \cite{nisan-91-1} that for \textit{monotone}
Boolean functions sensitivity and block sensitivity are equal.

A natural direction in the study of the gap between sensitivity
and block sensitivity is to restrict attention to Boolean
functions with symmetry. We note that a slight modification of
Rubinstein's construction (Example 2.13) gives a Boolean function,
invariant under the cyclic shift of the variables, which still
shows the quadratic gap between sensitivity and block sensitivity.
Tur\'an pointed out \cite{turan-84} that for \textit{symmetric}
functions (functions invariant under all permutations of the
variables), block sensitivity is within a factor of two of
sensitivity. For any non-trivial \textit{graph property} (the $n =
\binom{V}{2}$  variables indicate the adjacency relation among the $V$
vertices), Tur\'an \cite{turan-84} proved that sensitivity is at
least $V$ = $\Theta(\sqrt{n})$ and therefore the gap is at most
quadratic. In the same paper he also asked the following question:
\newline
\newline
\textbf{\underline{Problem (Tur\'an, 1984)}:} \textit{ Does a
lower bound of similar order hold still if we generalize graph
properties to Boolean functions invariant under a transitive
group of permutations?}

\

In Section 3 we give a cyclically invariant function
with sensitivity $\Theta (n^{1/3})$. This example gives a
negative answer to Tur\'an's question.

Kenyon and Kutin \cite{kutin-kenyon-01} observed that for ``nice"
functions the product of 0-sensitivity and 1-sensitivity tends to
be linear in the input length. Whether this observation extends to
all ``nice" functions was given as a (vaguely stated) open problem in
that paper.  In Section 3 we also construct a cyclically invariant Boolean
function for which the product of 0-sensitivity and 1-sensitivity is
$\Theta(\sqrt{n})$. Thus our function also gives a counterexample to
Kenyon and Kutin's suggestion.

In Section 2.1 we define a natural class of Boolean functions
called the minterm-transitive functions. It contains our new
functions (that we give in Section 3). In Section 4 we prove that
for minterm-transitive functions sensitivity is $\Omega(n^{1/3})$
(where $n$ is the input size) and the product of 0-sensitivity and
1-sensitivity is $\Omega(\sqrt{n})$. Thus for this class of
functions sensitivity and block sensitivity are polynomially
related.

\section{Preliminaries}

\subsection{Definitions}

We use the notation $[n] = \{1,2,3,...,n\}$. Let $f:\bool^n
\rightarrow \bool$ be a Boolean function. We call the elements of
$\bool^n$ ``words.'' For any word $x$ and $1\leq i \leq n$ we
denote by $x^i$ the word obtained by switching the $i$th bit of
$x$. For a word $x$ and $A \subseteq [n]$ we use $x^A$ to denote
the word obtained from $x$ by switching all the bits in $A$. For a
word $x = x_1,x_2,...,x_n$ we define $\supp(x)$ as $\{i \,|\, x_i
= 1\}$. Weight of $x$, denoted $\wt(x)$, is $|\supp(x)|$, $\ie$,
number of 1s in $x$.

\begin{definition}\rm{The \textit{sensitivity} of $f$  on the word $x$ is
defines as the number of bits on which the function is sensitive:
$s(f,x) = |\{i : f(x) \neq f(x^i)\}|$.

We define the \textit{sensitivity} of $f$ as $s(f)=\max\{s(f,x):x\in \bool^n\}$

We define \textit{0-sensitivity} of $f$ as
$s^{0}(f) = \max\{s(f,x):x\in \bool^n, f(x)=0\}$

We define \textit{1-sensitivity} of $f$ as
$s^{1}(f) = \max\{s(f,x):x\in \bool^n, f(x)=1\}$.
}
\end{definition}

\begin{definition}\rm{
The \textit{block sensitivity} $bs(f,x)$ of a function $f$ on an input $x$
is the maximum number of disjoint subsets $B_1, B_2, ..., B_r$ of
$[n]$ such that for all $j$, $f(x) \neq f(x^{B_j})$.

The \textit{block sensitivity} of $f$, denoted $bs(f)$, is $\max_x
bs(f,x)$.
}
\end{definition}

\begin{definition}\rm{
A \textit{partial assignment} is a function $p:S \to \bool$ where
$S \subseteq [n]$.  We call $S$ the support of this partial
assignment. The weight of a partial assignment is the number of
elements in $S$ that is mapped to 1. We call $x$ a (full)
assignment if $x:[n]\rightarrow \bool$. (Note than any word $x \in \bool^n$
can be thought of as a full assignment.) We say $p \subseteq x$ if
$x$ is an extension of $p$, $\ie$, the restriction of $x$ to $S$ denoted
$x|_S = p$.}
\end{definition}

\begin{definition}\rm{
A \textit{1-certificate} is a partial assignment, $p:S\rightarrow
\bool$, which forces the value of the function to 1. Thus if $x|_S
= p$ then $f(x)=1$.}
\end{definition}

\begin{definition}\rm{
If $\mathcal{F}$ is a set of partial assignments then we define
$m_{\mathcal{F}}:\bool^n \to \bool$ as $m_{\mathcal{F}}(x) = 1
\iff (\exists p\in \mathcal{F})$ such that $(p\subseteq x)$. }
\end{definition}

Note that each member of $\mathcal{F}$ is a 1-certificate for
$m_{\mathcal{F}}$ and $m_{\mathcal{F}}$ is the unique smallest such
function. (Here the ordering is pointwise, $\ie$, $f \leq g$ if for
all $x$ we have $f(x) \leq g(x)$).

\begin{definition}\rm{
A $minterm$ is a minimal 1-certificate, that is, no sub-assignment
   is a 1-certificate.}
\end{definition}

\begin{definition}\rm{ Let $S \subseteq$ [$n$] and let $\pi \in S_n$. Then
we define $S^{\pi}$ to be $\{ \pi(i) \,|\, i \in S\}$.

Let $G$ be a permutation group
acting on [$n$]. Then the sets $S^{\pi}$, where $\pi \in G$, are called the
\textit{G-shifts} of $S$.
If $p:S \rightarrow \bool$ is a partial assignment then we define
$p^{\pi}:S^{\pi} \rightarrow \bool$ as $p^{\pi} (i) = p(\pi^{-1}i)$.}
\end{definition}

\begin{definition} \rm{
Let $G$ be a subgroup of $S_n$, $\ie$, a permutation group acting on $[n]$.
A function $f:\bool^n \rightarrow \bool$ is said to be \textit{invariant under
the group} $G$ if for all permutations $\pi \in G$ we have $f(x^{\pi}) = f(x)$
for all $x \in \bool^n$.}
\end{definition}

\begin{definition}\rm{
Let $x = x_1x_2...x_n$ $\in \bool^n$  be a word. Then for $0<\ell<n$, we denote
by $cs_{\ell}(x)$ the word
$x_{\ell+1}x_{\ell+2}...x_nx_1x_2...x_{\ell}$, $\ie$, the \textit{cyclic
shift} of the variables of $x$ by $\ell$ positions.  }
\end{definition}

\begin{definition}\rm{
A function $f:\bool^n \rightarrow \bool$ is called
\textit{cyclically invariant} if $f(x) = f(cs_1(x))$
for all $x \in \bool^n$ .}
\end{definition}

Note that a cyclically invariant function is invariant under the group of
cyclic shifts.

\begin{proposition} \label{pro} Let $G$ be a permutation group.
Let $p:S\to \bool$ be a partial assignment and let
$\mathcal{F} = \{p^{\pi}\,|\, \pi \in G\}$.
Then $p$ is a minterm for the function $m_{\mathcal{F}}$.
\end{proposition}

The function $m_{\mathcal{F}}$ will be denoted $p^G$. Note that the function
$p^G$ is invariant under the group $G$. When $G$ is the group
of cyclic shifts we denote the function $p^{cyc}$. The function $p^{cyc}$
is cyclically invariant.

\

\begin{propproof} If $p$ has $k$ zeros then for any word $x$ with
fewer than $k$ zeros $m_{\mathcal{F}}(x) = 0$, since all the
element of $\mathcal{F}$ has same number of 1s and 0s. But if $q$
is a 1-certificate  with fewer than $k$ zeros we can have a word
$x$ by extending $q$ to a full assignment by filling the rest with
1s, satisfying $f(x)=1$ (since $q \subseteq x$). But $x$ contains
fewer than $k$ zeros, a contradiction. So no minterm of
$m_{\mathcal{F}}$ has fewer than $k$ zeros.

Similarly no  minterm of $\mathcal{F}$ has weight less than $p$.
So no proper sub-assignment of $p$ can be a 1-certificate. Hence
$p$ is a minterm of $m_{\mathcal{F}}$.
\end{propproof}

\

\begin{definition}\rm{
Let $G$ be a permutation group on $[n]$. $G$ is called
\textit{transitive} if for all
$1\leq i,j \leq n$ there exists a $\pi \in G$ such that $\pi(i) = j$.}
\end{definition}

\begin{definition}\rm{
Let $C(n,k)$ be the set of Boolean functions $f$ on $n$ variables
such that there exists a partial assignment $p:S\to \bool$ with
support $k(\neq 0)$ for which $f = p^{cyc}$. Let $C(n) = \cup
_{k=1}^nC(n,k)$. We will call the functions in $C(n)$
\textbf{minterm-cyclic}. These are the simplest cyclically
invariant functions.}
\end{definition}

\begin{definition}\rm{
Let $G$ be a permutation group on $[n]$. We define $D_G(n,k)$ (for
$k \neq 0$) to be the set of Boolean functions $f$ on $n$
variables such that there exists a partial assignment $p:S\to
\bool$ with support $k$ for which $f = p^G$. We define $D_G(n)$
to be $\cup_{k=1}^n D_G(n,k)$. This is a class of simple
$G$-invariant Boolean functions. We define $D(n)$ to be $\cup_G
D_G(n)$ where $G$ ranges over all transitive groups. We call these
functions \textbf{minterm-transitive}. Note that the class of
minterm-cyclic functions is a subset of the class of
minterm-transitive functions.}
\end{definition}

\subsection{Previous Results}

The largest known gap between sensitivity and block sensitivity is
quadratic, given by Rubinstein \cite{rubinstein-95}. Although
Rubinstein's example is not cyclically invariant, the following
slight modification is cyclically invariant with a
similar gap between sensitivity and block sensitivity.

\begin{example}
Let $g:\bool^k\rightarrow\bool$ be such that $g(x)=1$ iff $x$
contains two consecutive ones and the rest of the bits are 0. In
function $f':\bool^{k^2}\rightarrow\bool$ the variables are
divided into groups $B_1,\ldots, B_k$ each containing $k$
variables. $f'(x)=g(B_1)\vee g(B_2)\vee \cdots \vee g(B_k)$. Using
$f'$ we define the function $f:\bool^{k^2}\rightarrow\bool$ as
$f(x)=1$ iff $f(x')=1$ for some $x'$ which is a cyclic shift of
$x$. The sensitivity of $f$ is $2k$ while the block sensitivity is
$\lfloor \frac{k^2}{2}\rfloor$.
\end{example}

Hans-Ulrich Simon \cite{h.u.simon-83} proved that for any function
$f$ we have $s(f) \geq (\frac{1}{2}\log n - \frac{1}{2}\log\log n
+ \frac{1}{2})$, where $n$ is the number of effective variables
(the $i$th variable is effective if there exist some word $x$ for
which $f(x) \neq f(x^i)$). This bound is tight. Although for
various restricted classes of functions better bounds are known.

Let $f : \bool^m \rightarrow \bool$ be a Boolean function that
takes as input the adjacency matrix of a graph $G$ and evaluates
to 1 iff the graph $G$ has a given property. So the input size $m$
is $\binom{|V|}{2}$ where $|V|$ is the number of vertices in the
graph $G$. Also $f(G) = f(H)$ whenever $G$ and $H$ are isomorphic
as graphs. Such a function $f$ is called a \textit{graph
property}. Gy\"orgy Tur\'an \cite{turan-84} proved that graph
properties have sensitivity $\Omega(\sqrt{m})$.

A function $f$ is called \textit{monotone} if $f(x) \leq f(y)$
whenever $\supp(x) \subseteq \supp(y)$. Nisan\cite{nisan-91-1}
pointed out that for monotone functions sensitivity and block
sensitivity are the same.

In the definition of block sensitivity (Definition 2.2) if we
restrict the block size to be at most $\ell$ then we obtain the
concept of $\ell$-block sensitivity of the function $f$, denoted
$s_{\ell}(f)$. In \cite{kutin-kenyon-01} Kutin and Kenyon
introduced this definition and proved that $bs_{\ell}(f) \leq
c_{\ell}s(f)^{\ell}$ where $c_{\ell}$ is a constant depending on
$\ell$.

\section{The new functions}

In this section we will construct a cyclically invariant Boolean
function which has sensitivity $\Theta(n^{1/3})$ and a cyclically
invariant function for which the product of 0-sensitivity and
1-sensitivity is $\Theta(\sqrt{n})$.

\begin{theorem} There is a cyclically invariant function,
$f:\bool^n \rightarrow \bool$, such that, $s(f) =
\Theta(n^{1/3})$.
\end{theorem}

\begin{theorem} There is a cyclically invariant function,
$f:\bool^n \rightarrow \bool$, such that, $s^0(f)s^1(f)
=\Theta(\sqrt{n})$.
\end{theorem}

For proving the above theorems we will first define an auxiliary
function $g$ on $k^2$ variables ($k^2 \leq n$). Then we use $g$ to
define our new minterm-cyclic function $f$ on $n$ variables. If we
set $k = \lfloor n^{2/3}\rfloor$, Theorem 3.1 will follow. Theorem
3.2 follows by setting $k = \lfloor \sqrt{n}\rfloor$.

\

\textbf{The auxiliary function}

We first define $g:{\{0,1\}^{k^2}}\rightarrow \{0,1\}$ where $k^2
\leq n$. We divide the input into $k$ blocks of size $k$ each. We
define $g$ by a regular expression.
\[g(z) = 1 \iff z \in \underbrace{110^{k -2}}_{k}(\underbrace{11111(0+1)^{k -5}}_{k})^{k-2}\underbrace{11111(0+1)^{k -8}111}_{k}\ \ \ \
\cdots (1)\]

In other words, let $z\in \bool^{k^2}$ and let $z = z_1z_2...z_k$,
where each $z_i \in \bool^{k}$ for all $1\leq i\leq k$, $\ie$, $z$
is broken up into $k$ blocks of size $k$ each. Then $g(z) = 1$ iff
$z_1 = (11\underbrace{00...0}_{k-2})$ and for all $2\leq j \leq k$
the first five bits of $z_j$ are 1 and also the last 3 bits of
$z_k$ are 1. Note that $g$ does not depend on the rest of the
bits.

\

\textbf{The new function}

Now we define the function $f$ using the auxiliary function
$g$. Let $x|_{[m]}$ denote the word formed by the first $m$ bits
of $x$. Let us set
$$f(x)=1 \iff \exists \ell\ \textrm{such that}\ g\left(cs_{\ell}(x)|_{[k^2]}\right) =1.$$

In other words, viewing $x$ as laid out on a cycle, $f(x) = 1$ iff
$x$ contains a contiguous substring $y$ of length $k^2$ on which
$g(y) = 1$.

In other words, let $z\in \bool^{k^2}$ and let $z = z_1z_2...z_k$,
where each $z_i \in \bool^{k}$ for all $1\leq i\leq k$, $\ie$, $z$
is broken up into $k$ blocks of size $k$ each. Then $g(z) = 1$ iff
$z_1 = (11\underbrace{00...0}_{k-2})$ and for all $2\leq j \leq k$
the first five bits of $z_j$ are 1 and also the last 3 bits of
$z_k$ are 1. Note that $g$ does not depend on the rest of the
bits.

\

\textbf{Properties of the new function}

It follows directly from the definition that $f$
is a cyclically invariant Boolean function.

It is important to note that the function $g$ is so defined that
the value of $g$ on input $z$ depends only on $(6k-2)$ bits of
$z$.

Also note that the pattern defining $g$ is so chosen that if $g(z)
= 1$ then there is exactly one set of consecutive ($k - 2$) zeros in $z$ and
no other set of consecutive $(k-4)$ zeros.

\

\begin{claim} The function $f$ has
(a) 0-sensitivity $\Theta(\frac{n}{k^2})$ and (b) 1-sensitivity
$\Theta(k)$.
\end{claim}

\begin{clmproof} (a) Let $x$ be a word such that the first $k^2$
bits are of the form (1) and the rest of the bits are 0. Now
clearly $f(x) = 1$. Also it is easy to see that on this input $x$
1-sensitivity of $f$ is $(6k-2)$ and therefore $s^1(f) = \Omega(k)$.

Now let $x \in \bool^n$ be such that $f(x) = 1$ and there exists
$1\leq i\leq n$ such that $f(x^i) = 0$. But $f(x) = 1$ implies
that some cyclic shift of $x$ contains a contiguous substring $z$
of length $k^2$ of the form (1) (\ie, $g(z) = 1$). But since $g$
depends only on the values of $(6k-2)$ positions so one of those bits has to be
switched so that $f$ evaluates to 0. Thus $s^{1}(f) = O(k)$.

Combined with the lower bound $s^1(f) = \Omega(k)$ we conclude
$s^1(f) = \Theta(k)$.

(b) Let $\lfloor\frac{n}{k^2}\rfloor = m$ and $r = (n-k^2m)$. Let
$x =
(1\underline{0}0^{k-2}(111110^{k-5})^{k-2}111110^{k-8}111)^m0^r$.
Then $f(x) = 0$ since no partial assignment of the form (1) exists in
$x$. But if we switch any of the underlined zero the function
evaluates to 1. Note that the function is not sensitive on any
other bit. So on this input $x$ the 0-sensitivity of $f$ is $m =
\lfloor\frac{n}{k^2}\rfloor$ and therefore $s^0(f) = \Omega(\frac{n}{k^2})$.

Now let $x \in \bool^n$ and assume $f(x) = 0$ while $f(x^i)=1$ for
some $1\leq i\leq n$. By definition, the 0-sensitivity of $f$ is
the number of such values of $i$. For each such $i$ there exists a
partial assignment $z_i \subseteq x^i$ of the form (1). So $z_i^i$
is a contiguous substring of $x^i$ (or some cyclic shift of $x^i$)
of length $k^2$. Now consider the $z_i^i \subseteq x$ (recall
$z_i^i$ denotes the partial assignment obtained by switching the
$i$th bit of $z_i$). Due to the structure of the pattern (1) $z_i$
has exactly one set of consecutive $(k-2)$ zeros. So $z_i^i$ has
exactly one set of consecutive $(k-2)$ bits with at most one of
the bits being 1 while the remaining bits are zero. So the
supports of any two $z_i^i$ either have at least $(k^2-1)$
positions in common or they have at most two positions in common
(since the pattern (1) begins and ends with 11). Hence the number
of distinct $z_i^i$ is at most $\Theta(\frac{n}{k^2})$. Hence we
have $s^{0}(f) = O(\frac{n}{k^2})$.

Combined with the lower bound $s^0(f) = \Omega(\frac{n}{k^2})$ we conclude
that $s^0(f) = \Theta(\frac{n}{k^2})$.
\end{clmproof}
\newline
\newline
\textbf{Proof of Theorem 3.1:} From Claim 3.3 it follows $s(f) =
\max \left\{ \Theta(k), \Theta(\frac{n}{k^2})\right\}$ (since
$s(f) = \max{s^0(f), s^1(f)}$). So if we set $k = \lfloor
n^{2/3}\rfloor$ we obtain $s(f) = \Theta(n^{1/3})$.
{\hfill\rule{2mm}{2mm}}
\newline
\newline
\textbf{Proof of Theorem 3,2:} From Claim 3.3 we obtain
$s^0(f)s^1(f) = \Theta(\frac{n}{k})$. So if we set $k =
\lfloor\sqrt{n}\rfloor$ we have $s^0(f)s^1(f) =
\Theta(\sqrt{n})$. {\hfill\rule{2mm}{2mm}}

\

Theorem 3.1 answers Tur\'an's problem \cite{turan-84} (see the
Introduction) in the negative. In \cite{kutin-kenyon-01}, Kenyon and Kutin
asked whether $s^{0}(f)s^{1}(f) = \Omega(n)$ holds
for all ``nice" functions $f$. Although they do not define  ``nice,''
arguably our function in Theorem
3.2 is nice enough to answer the Kenyon-Kutin question is the
negative.

In the next section we prove that for a minterm-transitive
function, sensitivity is $\Omega(n^{1/3})$ and the product of
0-sensitivity and 1-sensitivity is $\Omega(\sqrt{n})$. Hence our
examples are tight.

\section{Minterm-transitive functions have sensitivity $\Omega(n^{1/3})$}

\begin{theorem}\label{thm1}
If $f$ is a minterm-transitive function on $n$ variables then
$s(f)=\Omega(n^{1/3})$ and $s^0(f)s^1(f) = \Omega(\sqrt{n})$.
\end{theorem}

To prove this theorem we will use the following three lemmas.
Since $f$ is a minterm-transitive function, $\ie$, $f \in D(n)$, we can say
$f \in D_G(n,k)$ for some transitive
group $G$ and some $k \neq 0$.

\begin{lemma}
If $f\in D_G(n,k)$ then $s^1(f) \ge \frac{k}{2}$.
\end{lemma}
\begin{proof}
Let $y$ be the minterm defining $f$. Without loss of
generality $\wt(y)\ge \frac{k}{2}$. Let us extend $y$ to a full
assignment $x$ by assigning zeros everywhere outside the support
of $y$.   Then switching any 1 to 0 changes the value of the
function from 1 to 0. So we obtain $s(f,x) \ge \frac{k}{2}$. Hence
$s^1(f) \ge \frac{k}{2}$.
\end{proof}

\begin{lemma}
If $S$ is a subset of $[n]$, $|S|=k$ then there exist at least
$\frac{n}{k^2}$ disjoint $G$-shifts of $S$.
\end{lemma}
\begin{proof}
Let $T$ be a maximal union of $G$-shifts of $S$. Since $T$ is
maximal $T$ intersects with all $G$-shifts of $S$. So we must have
$|T| \ge \frac{n}{k}$. So $T$ must be a union of at least
$\frac{n}{k^2}$ disjoint $G$-shifts of $S$. And this proves the
lemma.
\end{proof}

\begin{lemma}
If $f\in D_G(n,k)$ then $s^0(f) = \Omega(\frac{n}{k^2})$.
\end{lemma}
\begin{proof}
Let $y$ be the minterm defining $f$. By Lemma 2 we can have
$\Omega(\frac{n}{k^2})$ disjoint $G$-shifts of $y$. The union of
these disjoint $G$-shifts of $y$ defines a partial assignment. Let
$S=\{s_1, s_2, ..., s_r\}$ be the support of the partial
assignment. And let $Y_{s_i}$ be the value of the partial
assignment in the $s_i$-th entry.

Since $k \neq 0$ the function $f$ is not a constant function. Thus
there exists a word $z$ such that $f(z) = 0$. The $i$-th bit of
$z$ is denoted by $z_i$. We define,
\[T = \{j \,|\, z_j \neq  Y_{s_m}, s_m = j\}\]
Now let $P\subseteq T$ be a maximal subset of $T$ such that
$f(z^P)=0$. Since $P$ is maximal, if we switch any other bit in
$T\backslash P$ the value of the function $f$ will change to $1$.

So $s(f, z^P) \geq |(T \backslash P)|$. Now since $f(z^P)=0$ we
note that $z^P$ does not contain any $G$-shift of $y$. But from Lemma 4.3
we know that $z^T$
contains $\Omega(\frac{n}{k^2})$ disjoint $G$-shifts of $y$. So
$|(T\backslash P)|$ is $\Omega(\frac{n}{k^2})$ and thus $s^0(f)\geq s(f,z^P) =
\Omega(\frac{n}{k^2})$.
\end{proof}

\

\begin{thmproof1}
From the Lemma 4.2 and Lemma 4.4 we obtain,
$$s(f) = \max\{s^0(f), s^1(f)\} =
\max\left\{\Omega\left(\frac{n}{k^2}\right),\frac{k}{2}\right\}.$$
This implies $s(f) = \Omega(n^{1/3})$.

Now since $s^0(f)$ and $s^1(f)$ cannot be smaller than 1,
it follows from the Lemma 4.2 and 4.4 that
$$s^0(f)s^1(f) = \max\left\{\Omega\left(\frac{n}{k}\right), \frac{k}{2}\right\}.$$
So $s^0(f)s^1(f) = \Omega(\sqrt{n})$.
\end{thmproof1}

\

The new function we looked at in Theorem 3.1 is
minterm-transitive and has sensitivity $\Theta(n^{\frac
13})$. Thus this lower bound on sensitivity is tight for
minterm-transitive functions. Similarly for the function in
Theorem 3.2 the product of 0-sensitivity and 1-sensitivity is
tight.

An obvious corollary to the above theorem is,

\begin{corollary} If $f$ is minterm-transitive then $bs(f) = O(s(f)^3)$.
\end{corollary}

Hence for minterm-transitive functions, sensitivity and block
sensitivity are polynomially related.

\section{Open Problems}

The main question in this field is still open: Are sensitivity and
block sensitivity polynomially related? Can the gap between them
be more than quadratic? In fact we don't even know whether for all
minterm-transitive functions $f$ we have $bs(f) = O(s(f)^2)$ (that
is whether quadratic gap is the best possible gap even for
functions which are minterm-transitive). The following variant of
Tur\'an's question remains open:
\newline
\textbf{Problem:} If $f$ is a Boolean function invariant under a
transitive group of permutations then is it true that $s(f) \geq
n^c$ for some constant $c > 0$?

\textbf{\Large{\\Acknowledgements}}\\

I thank \,L\'aszl\'o Babai  and Sandy Kutin for giving me lots of
useful ideas and suggestions. I also thank Nanda Raghunathan for
helpful discussions. Sandy Kutin helped me simplify the proof of
Lemma 4.4.

\end{document}